\documentclass[12pt]{article}
\usepackage{epsf}
\usepackage{a4wide,graphicx}
\usepackage{cite}
\newcommand\T{\rule{0pt}{2.8ex}}
\newcommand\B{\rule[-1.2ex]{0pt}{0pt}}

\newcommand{\eps}{\epsilon}

\begin{document}

\title{Radiative open charm decay of the Y(3940), Z(3930), X(4160) resonances.}

\author{Wei Hong Liang$^1$, R. Molina$^2$ and E. Oset$^2$ \\
{\small{\it $^1$Physics Department, Guangxi Normal University, Guilin 541004, China}}\\
{\small{\it $^1$Departamento de F\'{\i}sica Te\'orica and IFIC,
Centro Mixto Universidad de Valencia-CSIC,}}\\
{\small{\it Institutos de
Investigaci\'on de Paterna, Aptdo. 22085, 46071 Valencia, Spain}}\\
}

\date{\today}

\maketitle

 \begin{abstract}

We determine the radiative decay amplitudes for decay into $D^*$ and $\bar{D}
\gamma$, or $D^*_s$ and $\bar{D}_s \gamma$ of some of the charmonium like states
classified as X,Y,Z resonances, plus some other hidden charm states which are
dynamically generated from the interaction of vector mesons with charm. The
mass distributions as a function of the $\bar{D} \gamma$ or $\bar{D}_s \gamma$
invariant mass show a peculiar behavior as a consequence of the $D^* \bar{D}^*$
nature of these states. The experimental search of these magnitudes can shed
light on the nature of these states.

\end{abstract}

\section{Introduction}

  The use of the chiral unitary approach, combining chiral dynamics and
unitarity in coupled channels has allowed one to study the interaction of
pseudoscalar mesons, and of pseudoscalar mesons with baryons, at higher 
energies than allowed by
perturbation theory \cite{review}. One of the peculiar findings of the approach
is that many resonances appear as poles in the scattering matrix as a
consequence of the interaction, which are called dynamically generated
resonances, and account for many of the low lying scalar meson and axial vector
states, as well as for the low lying baryonic resonances.  The success of this
theory, providing properties of these states and accurate cross sections in
production reactions, has stimulated the extension to the interaction of
vector mesons.

A natural extension of the chiral Lagrangians to incorporate vector mesons and
their interaction is provided by the hidden local gauge formalism for vector
interactions with pseudoscalar
mesons, vectors and photons \cite{hidden1,hidden2,hidden3,Bernard:1988db},
which provides a consistent and
successful scheme to address many issues of hadron physics. Yet, as was the case
with the interaction of pseudoscalar mesons or pseudoscalar mesons with baryons,
also here it is the combination of the interaction provided
by these Lagrangians with unitary techniques in coupled channels that allows one
to obtain a realistic approach to the vector-vector interaction.
  In this direction, the work of
\cite{raquel,geng} has allowed to study the vector-vector interaction at
intermediate energies, up to about 2000 MeV, where the nonperturbative unitary
techniques are essential since many resonances are generated as a consequence
of the interaction. In practice one solves a set of coupled channels Bethe
Salpeter equations using as kernel the interaction provided by the hidden gauge
Lagrangians and regularizing loops with a natural scale \cite{ollerulf}. The
results of \cite{raquel} show that the $f_0(1370)$ and $f_2(1270)$ mesons are
dynamically generated from the $\rho \rho$ interaction. Actually, there are
strong experimental arguments to suggest that the  $f_0(1370)$ is a $\rho \rho$
molecule \cite{klempt,crede}.

The work of
\cite{raquel} has been extended to the interaction of all members of the vector
nonet with the result that eleven resonances are dynamically generated,
most of which can be associated
to known resonances, while other ones remain as predictions \cite{geng}.

 Some predictions of this approach for physical processes involving these
 states have readily followed to further support their
nature as dynamically generated. In this sense
 the radiative decay of the
$f_0(1370)$ and $f_2(1270)$ mesons into $\gamma \gamma$ \cite{yamagata}, were
found in agreement with the experimental data. Similarly the $J/\psi$ decay into
$\phi (\omega)$ and one of the $f_2(1270)$, $f'_2(1525)$, $f_0(1710)$ resonances,
and into $K^*$ and the $K^*_2(1430)$  \cite{chinacola}, was also found consistent
with experiment. In the same line, the  $J/\psi$ radiative decay into
$\gamma$ and one of these nonstrange resonances was also found in agreement with
experimental data \cite{chinavalgerman}. More recently, the work of \cite{yamagata} has been extended in \cite{BranzGeng} to study the $\gamma\gamma$ and $\gamma$-vector meson decays of the eleven dynamically generated resonances of \cite{geng}, with also good agreement with experiment in the cases that there are data.

In \cite{raquelnaga}
 the extension to include charm mesons has been done studying the interaction of the $\rho$ meson with $D^*$
 mesons, where three states are obtained, one which can be easily associated
 with the tensor state $D^*_2(2460)$, another one which is very likely to be
 the $D^*(2640)$ in view of its mass and the natural
 explanation for the small width compared with that from $D^*_2(2460)$, and a third one which corresponds to
 a scalar meson, for which no counterpart is yet reported in the PDG \cite{pdg}.

 More recently the work has been extended to the interaction of $D^* \bar{D}^*$
in \cite{xyz}, where five resonances are dynamically generated, three of which
could be tentatively associated to some X, Y, Z resonances reported recently,
concretely the  Y(3940), Z(3930), X(4160).

Independently, an alternative approach to the hidden gauge formalism,
based on chiral symmetry and heavy quark symmetry, but only in one channel, has
been used in \cite{shilinzhu}, where also bound states of the
$D^* \bar{D}^*$ systems are found in some cases.

   Following the idea of \cite{shilinzhu,gutsche} that a Y(3930) and the Y(4140) in \cite{xyz}
 could be actually $D^* \bar{D}^*$ and $D^*_s \bar{D}^*_s$ molecules,
 respectively, an idea was given in \cite{liuke} that the shape of the spectrum
 in the radiative
 decay of these resonances into  $D^*\bar{D}\gamma$, or $D^*_s\bar{D}_s\gamma$,
 respectively, can further test the molecular assignment of these two resonances.

   We follow the idea of  \cite{liuke} with a different technical approach, and
 for the five states dynamically generated in \cite{xyz}.  The work of
 \cite{xyz} provides scattering amplitudes for  $D^*\bar{D}^*$ and its coupled
 channels. From there, evaluating the residues at the poles, one determines the
 coupling of the resonances to the different channels and this is all that one
 needs to evaluate the radiative decay in the Feynman diagrammatic approach
 that we follow. This allows us to determine not only shapes of the spectra but
 also absolute numbers for the radiative decay in terms of the $D^* D
 \gamma$ coupling that can be taken from the experiment in the case of $D^{*+}\to D^+ \gamma$ (and  $D^{*-}\to D^- \gamma$), and ratios of the $R\to D^*\bar{D}\gamma$ decay width to the radiative width of the $D^*(D^*_s)$ states in general. 
 In \cite{liuke} a different
 method based on wave functions of the states is reported and no absolute values
 are provided. In addition we give arguments on why the $X\to D^*\bar{D}\gamma$ distribution with
 respect to the $\bar{D} \gamma$ is the observable which connects easier with the
 dynamically generated nature of the resonances (molecular nature in the wave
 function picture). 

\section{Formalism}
In \cite{xyz} a coupled channel formalism was considered in which one had essentially the
hidden charm $D^* \bar{D}^*$, $D^*_s \bar{D}^*_s$ pairs plus all the
charmless vector-vector pairs like $\rho \rho$, $\rho \omega$  
or $\phi \phi$, which have the same quantum numbers of the states that are
investigated and provide the decay width of the XYZ states obtained. Five heavy states were generated, additional to the light ones
obtained from the light vector pairs in \cite{geng}, three of which were
identified with
states observed at Belle and Babar, the Y(3940), Z(3930) and X(4160), and two other
states, so far not observed, which were called $Y_p(3945)$ and $Y_p(3912)$.  The
quantum numbers of the states and their assumed  experimental counterparts are
summarized in Table \ref{tab:exp}.

\begin{table}[htb!]
\begin{center}
\begin{tabular}{c|c|c|c|c|c|c}
$I^G[J^{PC}]$&\multicolumn{2}{c|}{Theory}&\multicolumn{4}{c}{Experiment}\\
\hline\hline
& Mass [MeV] & Width [MeV]\T\B & Name & Mass [MeV] & Width [MeV] &$J^{PC}$\\
$0^+[0^{++}]$&$3943$&$17$&Y(3940)&$3943\pm 17$&$87\pm 34$&$J^{P+}$\\
& & & & $3914.3^{+4.1}_{-3.8}$ & $33^{+12}_{-8}$ & \\
$0^-[1^{+-}]$&$3945$&$0$&"$Y_p(3945)$"& & & \\
$0^+[2^{++}]$&$3922$&$55$&Z(3930)&$3929\pm 5$& $29\pm 10$&$2^{++}$\\
$0^+[2^{++}]$&$4157$&$102$&X(4160)&$4156\pm 29$& $139^{+113}_{-65}$& $J^{P+}$\\
$1^-[2^{++}]$&$3912$&$120$&"$Y_p(3912)$"& & & \\
\hline
\end{tabular}
\end{center}
\caption{Comparison of the mass, width and quantum numbers with the experiment.}
\label{tab:exp}
\end{table}

\begin{table}[htb]
\begin{center}
\begin{tabular}{c|cccc}
 & Y(3940)&$Y_p(3945)$&$Z(3930)$&$X(4160)$\T\B\\
 \hline
 \hline
 channel& \multicolumn{4}{c}{$|g_i|$ (MeV)\T\B}\\
\hline
$D^* \bar{D}^*$&$18822$&$18489$&$21177$\T\B&$1319$\\
\hline
$D^*_s \bar{D}^*_s$&$8645$&$8763$&$6990$&$19717$\T\B\\
\hline
$K^* \bar{K}^*$&$15$&$40$&$44$\T\B&$87$\\
\hline
$\rho \rho$&$52$&$0$&$84$&$73$\T\B\\
\hline
$\omega \omega$&$1368$&$0$&$2397$\T\B&$2441$\\
\hline
$\phi \phi$&$1011$&$0$&$1999$&$3130$\T\B\\
\hline
$J/\psi J/\psi$&$422$&$0$&$1794$&$2841$\T\B\\
\hline
$\omega J/\psi$&$1445$&$0$&$3433$\T\B&$2885$\\
\hline
$\phi J/\psi$&$910$&$0$&$3062$&$5778$\T\B\\
\hline
$\omega\phi$&$240$&$0$&$789$&$1828$\T\B\\
\hline
\end{tabular}
\caption{Modules of the coupling in the $I=0,J=0,1,2$ sectors.}
\label{tabc1}
\end{center}
\end{table}

\begin{figure}
\centering
\includegraphics[width=6cm]{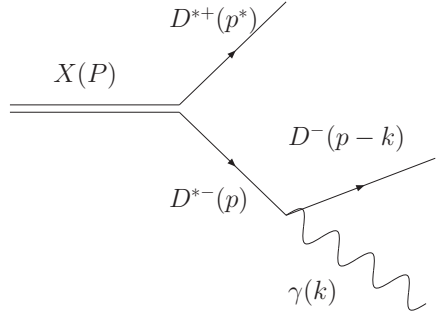} \\
\caption{Decay of the X resonance to $D^{*+}D^-\gamma$.}
\label{fig:fig1}
\end{figure}

\begin{table}[htb]
\begin{center}
\begin{tabular}{c|c}
 & $Y_p(3912)$ \T\B\\
\hline
\hline
 channel&$|g_i|$ (MeV)\T\B\\
 \hline
$D^* \bar{D}^*$&$20869$\T\B\\
\hline
$K^* \bar{K}^*$&$152$\T\B\\
\hline
$\rho\rho$&$0$\T\B\\
\hline
$\rho\omega$&$3656$\T\B\\
\hline
$\rho J/\psi$&$6338$\T\B\\
\hline
$\rho\phi$&$2731$\T\B\\
\hline
\end{tabular}
\caption{Modules of the coupling in the $I=1,J=2$ sector.}
\label{tabc2}
\end{center}
\end{table}
In \cite{xyz} the states were identified by observing poles in the
vector-vector scattering matrix with certain quantum numbers.  The real part of
the pole position provides the mass of the resonance and the imaginary part
half its width. In addition the residues at the poles provide the product of the
couplings of the resonance to the initial and final channels, from where, by
looking at the scattering amplitudes in different channels, we can obtain the
coupling of the resonance to all channels up to an irrelevant global sign, which is assigned
to one particular coupling. In Tables \ref{tabc1} and \ref{tabc2}, the couplings to the channels are
also shown. As one can see from these tables, the states obtained correspond to basically bound
$D^* \bar{D}^*$
or $D^*_s \bar{D}_s^*$ states, hence the decay into these pairs is forbidden, whereas the light vector-light vector channels provide the width of the states.
However, if one looks at the decay channel of the $\bar{D}^*$ into
$\bar{D}\gamma$, then the process $X \to D^* \bar{D} \gamma$ is allowed, since
the mass of the resonance X, for all the cases listed in Table \ref{tab:exp}, exceeds the
sum of masses of the final state. In Fig. \ref{fig:fig1} the corresponding Feynman diagram to the $X\to D^{*+}D^- \gamma$ process is shown. The  $D^{*-}$ propagates virtually between
the production point of $X \to D^{*+} D^{*-}$ and the decay point of
$D^{*-}\to D^-\gamma$. This propagator is the relevant characteristic of the $X\to D^{*+}D^- \gamma$ decay. Thus, this diagram is peculiar to the assumed nature of the resonance X as a
molecule of $D^* \bar{D}^*$ and should be largely dominant over other possible
processes \cite{liuke}.  The evaluation of this Feynman diagram is easy. All one
needs is the coupling of the resonance to $D^{*+} D^{*-}$, together with the
corresponding spin projection operator, and the vertex accounting for the decay
of $D^{*-}$ into $D^-\gamma$.

\begin{figure}
\centering
\includegraphics[width=6cm]{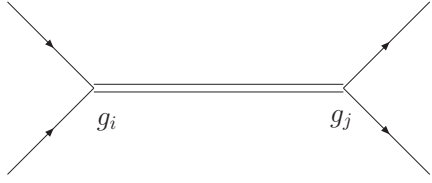} \\
\caption{Representation of the $T$ matrix obtained from the Bethe Salpeter Equation in \cite{raquelnaga}.}
\label{fig:fig2}
\end{figure}

 The spin projection operators on $J=0,1,2$, evaluated assuming the 
 three momenta of the $D^*$ and  $\bar{D}^*$
  to be small with respect to the mass of the charmed vector
 mesons, which is indeed the case here, are given in terms of the polarization
 vectors by
\begin{eqnarray}
{\cal P}^{(0)}&=& \frac{1}{3}\eps^{(1)}_i \eps^{(2)}_i \eps^{(3)}_j \eps^{(4)}_j\nonumber\\
{\cal P}^{(1)}&=&\frac{1}{2}(\eps_i^{(1)}\eps_j^{(2)}-\eps_j^{(1)}\eps_i^{(2)})\frac{1}{2}(\eps_i^{(3)}\eps_j^{(4)}-\eps_j^{(3)}\eps_i^{(4)})\nonumber\\
{\cal P}^{(2)}&=&\lbrace\frac{1}{2}(\eps_i^{(1)}\eps_j^{(2)}+\eps_j^{(1)}\eps_i^{(2)})-\frac{1}{3}\eps_l^{(1)}\eps_l^{(2)}\delta_{ij}\rbrace\nonumber\\
&\times& \lbrace\frac{1}{2}(\eps_i^{(3)}\eps_j^{(4)}+\eps_j^{(3)}\eps_i^{(4)})-\frac{1}{3}\eps_m^{(3)}\eps_m^{(4)}\delta_{ij}\rbrace \ .
\label{eq:projectores}
\end{eqnarray}
The amplitude obtained after summing all the diagrams included implicitly in the Bethe Salpeter equation, $T=[1-VG]^{-1}\,V$, goes close to a pole, as ${\cal P}^{(k)}\,g_i\, g_j/(s-s_p)$, where $g_{i(j)}$ is the coupling of the resonance to the $i(j)$ channel and ${\cal P}^{(k)}$ are the spin projectors over spin $k=0,1,2$ of Eq. (\ref{eq:projectores}), see \cite{raquel}. This final amplitude is depicted in the diagram of Fig. \ref{fig:fig2}. In this way, if we take the case of the Y(3940), with $J=0$, the first vertex in the diagram of Fig. \ref{fig:fig1} is $\frac{1}{\sqrt{3}}\eps^{(1)}_i \eps^{(2)}_i \,g_{D^*\bar{D}^*}\, \mathrm{F_I}$, where $\mathrm{F_I}$ is the isospin factor needed to change from the isospin basis, where the couplings are evaluated in \cite{xyz}, to the charge basis. In the case of $D^{*+}D^{*-}$, we have $\mathrm{F_I}=\frac{1}{\sqrt{2}}$. In what follows, we will call $\tilde{g}$ the coupling of the resonance to the $VV$ state in isospin basis.

 On the other hand the anomalous vertex for the $\bar{D}^*$ decay
 into $\bar{D} \gamma$ is given by
\begin{equation}
-it_{\bar D^* \to \bar D \gamma}=-ig_{PV\gamma
}\,\epsilon_{\mu\nu\alpha\beta}\,p^{\mu}\,\epsilon^{\nu}(\bar
D^*)\,k^{\alpha}\,\epsilon^{\beta}(\gamma),
\label{eq:anom}
\end{equation}
where $p$, $k$ are the momenta of the $D^{*-}$ and $\gamma$ respectively. This amplitude gives rise to a width
\begin{equation}
\Gamma_{\bar{ D}^* \to \bar{ D} \gamma}=\frac{1}{48\pi}g^2_{PV\gamma
}\frac{k}{M^2_{\bar{D}^*}}(M^2_{\bar{D}^*}-m^2_{\bar{D}})^2.
\label{eq:radwidth}
\end{equation}

 Unfortunately, only the value for the radiative decay of the
 $D^{*-} \to D^- \gamma$ and of its positive state partner are known. In this
 case we will be able to provide an absolute value for the radiative decay
 width of the XYZ resonances. In the other cases we will give the ratio of the
 radiative decay of the resonance to that of the $\bar{D}^*$.  The value of $g_{PV\gamma}$
 for the $D^{*-} \to D^- \gamma$ decay is given by
 \begin{equation}
 g_{PV\gamma}= 1.53 \times10^{-4} MeV^{-1},
\end{equation}
which can be easily deduced using Eq. (\ref{eq:radwidth}) from the experimental value
of the width $\Gamma= 1.54 $ KeV.

   Let us begin with the decay of the Y(3940). This state has isospin zero and
   spin zero. According to \cite{xyz} it couples mostly to $D^* \bar{D}^*$, has a
   smaller coupling to $D^*_s \bar{D}^*_s$ and very small coupling to pairs of
   charmless vectors, see Table \ref{tabc1}.  The couplings in \cite{xyz} are given in isospin basis.
   However, we need them now in charge basis, which are readily obtained for the
   isospin combinations
\begin{eqnarray}
|D^*\bar D^*,I=0, I_3=0\rangle &=&\frac{1}{\sqrt{2}}(D^{*+}D^{*-} +
D^{*0}\bar
D^{*0}),\nonumber \\
|D^*\bar D^*,I=1, I_3=0\rangle &=&\frac{1}{\sqrt{2}}(D^{*+}D^{*-} -
D^{*0}\bar
D^{*0}), \\
|D_s\bar D_s,I=0, I_3=0\rangle &=&D_s^{*+}D_s^{*-}.\nonumber
\end{eqnarray}
   Thus, the couplings of \cite{xyz} for $D^* \bar{D}^*$ must be multiplied by
   $1/\sqrt 2$ to get the appropriate coupling for the charged or neutral states
   (a sign is irrelevant for the width), and do not require an extra factor
    for the case of $D^*_s \bar{D}^*_s$.

   With the previous information we can already write the amplitude for the
   decay of the Y(3940) into $D^{*+}  D^- \gamma$, which is given by
\begin{eqnarray}
-it&=&-i\frac{1}{\sqrt{2}}\,\tilde{g}\,\frac{1}{\sqrt{3}}\epsilon_i^{(1)}\epsilon_i^{(2)}\,\frac{i}{p^2-M^2_{D^*}+iM_{D^*}\Gamma_{D^*}}\nonumber \\
&&\times (-i)\,g_{PV\gamma}\,\epsilon_{\mu\nu\alpha\beta}\,p^{\mu}\epsilon^{\nu
(2)}\,k^{\alpha}\,\epsilon^{\beta}(\gamma),
\end{eqnarray}
   where the indices (1), (2) indicate the $D^{*+}$ and the $D^{*-}$ respectively.
   The sum over the intermediate $D^{*-}$ polarizations can be readily done as
 \begin{equation}
 \sum\limits_{\lambda}\epsilon_i^{(2)}\epsilon^{\nu
 (2)}=-g_i^\nu=-\delta_{i\nu},
 \end{equation}
   where we have neglected the three momenta of the intermediate $D^{*-}$ which
   is in average very small compared with the $D^{*-}$ mass, particularly at
   large invariant masses of the $D^{-} \gamma$ system which concentrates most
   of the strength, as we shall see. The sum  of $|t|^2$ over the final
   polarizations of the vector and the photon is
   readily done and, neglecting again terms of order $\vec{p}\,^2/M_{D^*}^2$,
    we get the  result
\begin{eqnarray}
\sum|t|^2&=&\frac{1}{3}\frac{1}{2}\tilde{g}^2g^2_{PV\gamma}\left|\frac{1}{p^2-M^2_{D^*}+iM_{D^*}\Gamma_{D^*}}\right|^2
2(p\cdot k)^2\nonumber \\
 &=&\frac{1}{6}\frac{1}{2}\tilde{g}^2g^2_{PV\gamma}\left|\frac{p^2-m^2_D}{p^2-M^2_{D^*}+iM_{D^*}\Gamma_{D^*}}\right|^2.
\label{eq:tdos}
 \end{eqnarray}

   The differential mass distribution with respect to the  invariant mass of the
   $D^{-} \gamma$ system, $M_{inv}$, with $M_{inv}^2=p^2$, is finally given by
\begin{equation}
\frac{d
\Gamma_R}{dM_{inv}}=\frac{1}{4M_R^2}\frac{1}{(2\pi)^3}\,p^*
\tilde{p}_D\sum|t|^2,
\label{eq:dist}
\end{equation}
   where $p^*$ is the momentum of the $D^{*+}$ in the rest frame of the
   resonance X and $\tilde{p}_D$ is the momentum of the $D^-$ in the rest
   frame of the final  $D^{-} \gamma$ system given by
\begin{eqnarray}
p^*&=&\frac{\lambda^{1/2}(M_R^2,M^2_{D^*},M^2_{inv})}{2M_R},\nonumber \\
\tilde{p}_D&=&\frac{M^2_{inv}-m^2_D}{2M_{inv}}.
\end{eqnarray}

   In the case of the tensor and spin one states we must do extra work since the
projector operators are different. In this case we must keep the indices $i, j$
in $t$ and multiply with $t^*$ with the same indices $i, j$ and then perform the sum over the indices $i, j$. This sums over all
possible final polarizations but also the initial X polarizations, so in order
to take the sum and average over final and initial polarizations, respectively,
one must divide the results of the $\sum_{i,j}\,tt^*$ by $(2J+1)$, where $J$ is
the spin of the resonance X.  The explicit evaluation for the case of the tensor
states, $J=2$, of $D^* \bar{D}^*$ proceeds as follows: The $t$ matrix is now
written as
\begin{eqnarray}
t&=&\frac{1}{\sqrt{2}}\,\tilde{g}\,g_{PV\gamma} \,\left\{
\frac{1}{2}\left(\epsilon_i^{(1)}\epsilon_j^{(2)}+\epsilon_j^{(1)}\epsilon_i^{(2)}
\right)
-\frac{1}{3}\epsilon_l^{(1)}\epsilon_l^{(2)}\delta_{ij}\right\}\nonumber \\
&\times&\frac{1}{p^2-M^2_{D^*}+iM_{D^*}\Gamma_{D^*}}\,\epsilon_{\mu\nu\alpha\beta}\,p^{\mu}\epsilon^{\nu
(2)} k^\alpha\epsilon^\beta(\gamma).
\end{eqnarray}

 As mentioned above, we must multiply $t_{i,j}$ by $t^*_{i,j}$, recalling that
 the indices $i,j$ are spatial indices and divide by $(2J+1)$ (5 in this case)
 in order to obtain the modulus squared of the transition matrix, summed and
 averaged over the final and initial polarizations.  Neglecting again terms
 that go like $\vec{p}\,^2/m_D^{*2}$ we obtain the same expression as in 
 Eq. (\ref{eq:tdos}).
   It is also easy to see that this is again the case for the $J=1$ states.  The
 normalization of the spin projection operators in Eq. (\ref{eq:projectores})
   makes this magnitude to be the same in all cases.
\section{Convolution of the $d\Gamma/d M_{inv}$ due to the width of the XYZ states}

Some of the dynamically generated XYZ states have a non negiglible width and, as a consequence, a mass distribution. That means there is a probability of these states to have a mass over the nominal mass and if one consider this fact, the $PV\gamma$ decay width should increase. In order to consider this, we convolute the $d\Gamma/d M_{inv}$ function over the mass distribution of the resonance $R$. We take $\Gamma/2$ to both sides of the peak of the resonance distribution which account for a large fraction of the strength and produces distinct shapes in the $\gamma \bar{D}$ mass distribution. We find:

\begin{equation}
d\Gamma^{\mathrm{conv(\Gamma/2)}}/d M_{inv}=\frac{1}{N}\,\int^{(M_R+\Gamma/2)^2}_{(M_R-\Gamma/2)^2} d\tilde{M}^2\, (-\frac{1}{\pi})\, Im\frac{1}{\tilde{M}^2-M_R^2+i \Gamma M_R}\,d\Gamma/d M_{inv}
\label{eq:conv}
\end{equation}
with
\begin{equation}
N=\int^{(M_R+\Gamma/2)^2}_{(M_R-\Gamma/2)^2}d\tilde{M}^2\, (-\frac{1}{\pi})\, Im\frac{1}{\tilde{M}^2-M_R^2+i \Gamma M_R}\ ,\nonumber
\end{equation}
As we will see in the next section, the use of Eq. (\ref{eq:conv}) leads to an increase of $\Gamma(R\to PV\gamma)$ with respect to the result with the nominal mass $M_R$.
\section{Results}

We show here the results for different cases:

\subsection{The Y(3940): Decay mode $D^{*+}  D^- \gamma$}
 The results are the same reversing the signs of the charges.

\begin{figure}
\centering
\includegraphics[width=0.8\textwidth]{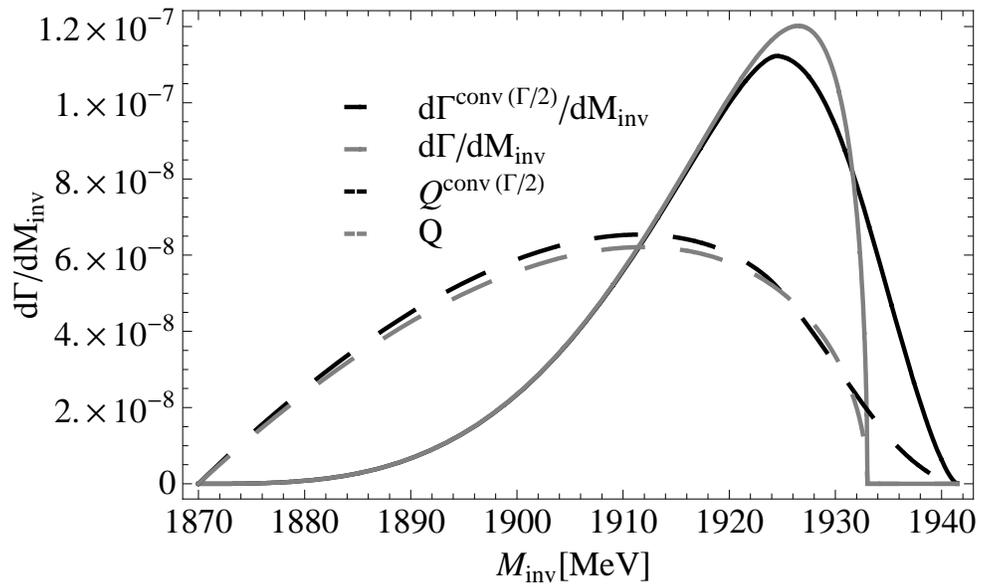}
\caption{The $Y(3940)\to D^{*+}D^-\gamma$: Comparison of $d\Gamma/dM_{inv}$ and $d\Gamma^{\mathrm{conv(\Gamma/2)}}/dM_{inv}$, $Q$ and $Q^{\mathrm{conv(\Gamma/2)}}$ as a function of the $D^-\gamma$ invariant mass.}
\label{fig:comp}
\end{figure}

   In Fig. \ref{fig:comp} we show the distribution of Eq. (\ref{eq:dist}), together with Eq. (\ref{eq:tdos}), between the limits of $M_{inv}$: $m_D$
   and $M_R -m_{D^*}$. Also, in order to see the effects produced when one considers the width of the state, we plot in the same figure $d\Gamma^{\mathrm{conv}}/dM_{inv}$, taken from Eq. (\ref{eq:conv}).
    We can see a very distinct picture, with most of the strength accumulated at
the maximum values of $M_{inv}$.  The propagator of the intermediate $D^{*-}$
and the factor $(p.k)^2$ are
responsible for that shape. In fact we show superposed in the same figure the
result obtained ( normalized to the same area) substituting the propagator by a
constant and removing the factor $(p.k)^2$ (or equivalently the factor
$(p^2-m_D^2)^2$). We call $Q$ the resulting distribution (or $Q^{\mathrm{conv}}$ when one convolute this function taking into account the width of the $R$ state).  We can see that the pictures of $d\Gamma/dM_{inv}$ and  $Q$ (or equivalently $d\Gamma^{\mathrm{conv}}/dM_{inv}$ and $Q^{\mathrm{conv}}$) are radically different
and the reason is mostly due to the presence of the $D^{*-}$ propagator which carries the
memory that the resonance Y(3940) is assumed to be a $D^* \bar{D}^*$ molecule. The effects of considering the convolution are also visible in this picture. Now, $d\Gamma^{\mathrm{conv}}/dM_{inv}$ spreads beyond $M_R-m_{D^*}$, and there is some probability for the state to decay into $PV\gamma$ up to $M_{inv}=M_R+\Gamma/2-m_{D^*}$, where $\Gamma$ is the width of the state. Also in this case, the difference between $d\Gamma^{\mathrm{conv}}/dM_{inv}$ and $Q^{\mathrm{conv}}$ is clearly visible.

For the case of decay into $D^{*0}  \bar{D}^0 \gamma$ the matrix element is
formally the same except that now we do not know
 the experimental radiative decay width of the $\bar{D}^{*0}$. In this case we
divide the mass distribution of the $D^{*0}  \bar{D}^0 \gamma$ decay by the
width of the $\bar{D}^{*0}\to  \bar{D}^0\gamma$ and plot the magnitude
\begin{eqnarray}
\frac{1}{\Gamma_{D^* \to D\gamma}}\frac{d
\Gamma_R}{dM_{inv}}&=&\frac{1}{2}\frac{1}{6}\,\tilde{g}^2g^2_{PV\gamma}
\left|\frac{p^2-m^2_D}{p^2-M^2_{D^*}+iM_{D^*}\Gamma_{D^*}}\right|^2 \nonumber \\
&&\times\frac{48\pi\,
M^2_{D^*}}{k(M^2_{D^*}-m^2_D)^2}\,\frac{1}{4M^2_R}\,\frac{1}{(2\pi)^3}\,p^*\tilde{p}_D,
\label{eq:gammarel}
\end{eqnarray}
with
$$
k=\frac{M^2_{D^*}-m^2_D}{2M_{D^*}}.
$$

\begin{figure}
\centering
\includegraphics[width=0.8\textwidth]{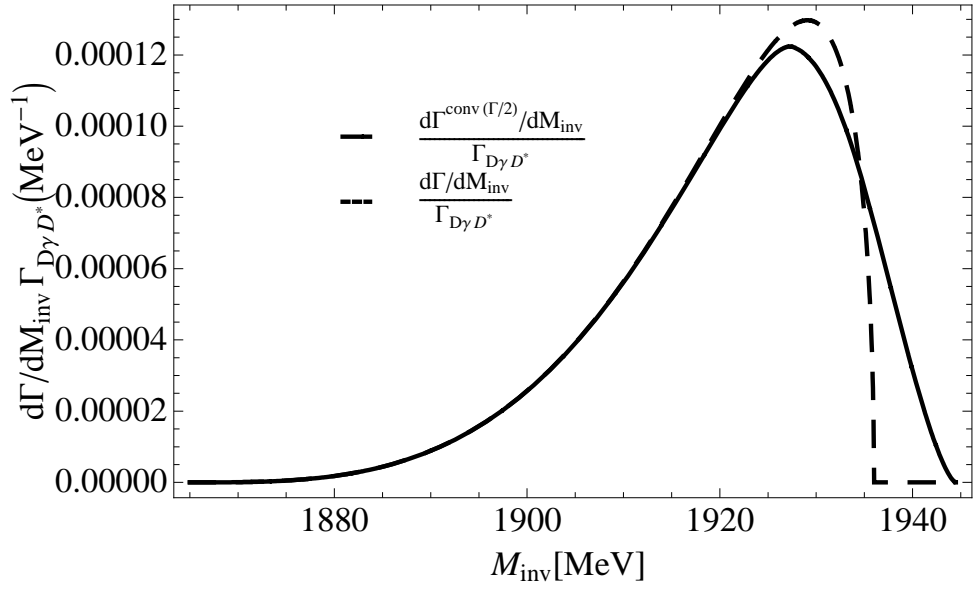}
\caption{The $Y(3940)\to D^{*0}\bar{D}^0\gamma$: Comparison of $d\Gamma/dM_{inv}\Gamma_{D^* D\gamma}$ and  $d\Gamma^{\mathrm{conv(\Gamma/2)}}/dM_{inv}\Gamma_{D^*D\gamma}$.}
\label{fig:comp1}
\end{figure}

 In Fig. \ref{fig:comp1} we show the results of the $d\Gamma_R/dM_{inv}\Gamma_{D^* D\gamma}$ distribution and also we compare with $d\Gamma^{\mathrm{conv}}_R/dM_{inv}\Gamma_{D^*D\gamma}$. We can see that the enlarged range of the mass distribution between the limits $M_{inv}=M_R-m_{D^*}$ and $M_R+\Gamma/2-m_{D^*}$ is responsible for an increase in $\Gamma(Y(3940)\to D^{*0}\bar{D}^0\gamma)$.

\subsection{The $Y_p(3945)$}
\begin{figure}
\centering
\includegraphics[width=0.8\textwidth]{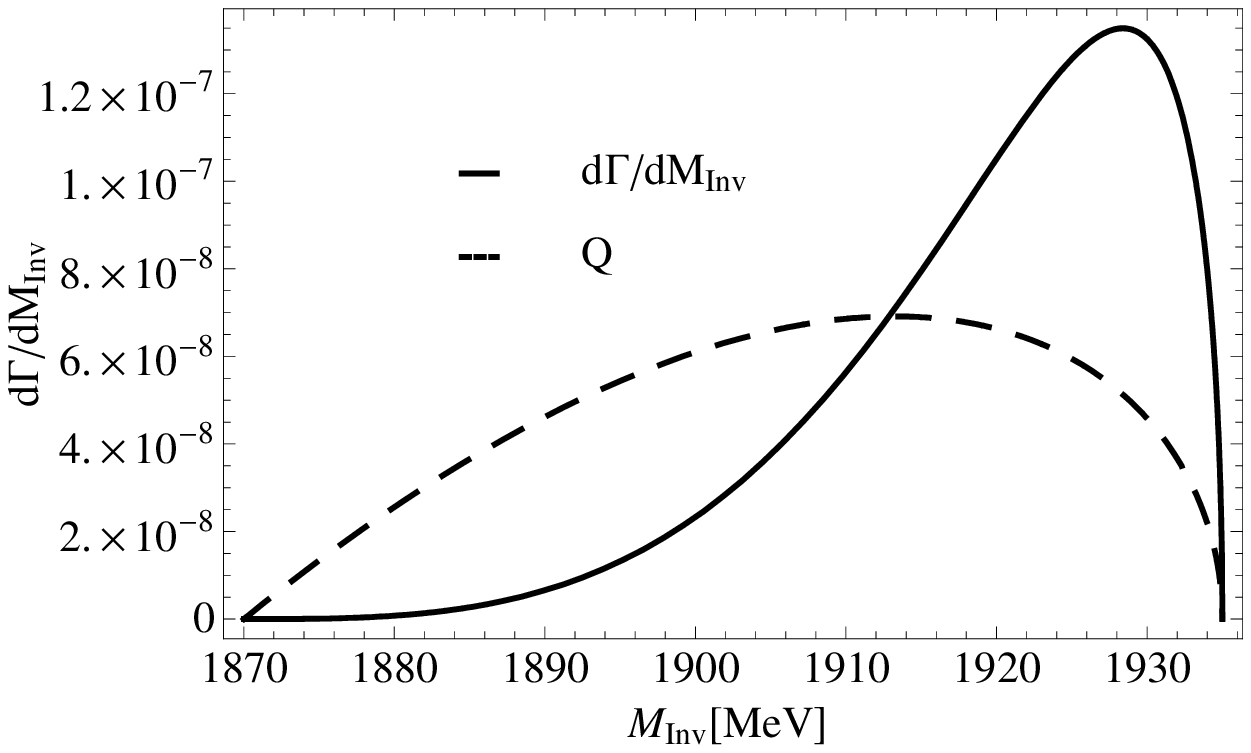}
\caption{The $Y_p(3945)\to D^{*+}D^-\gamma$: $d\Gamma/dM_{inv}$ and $Q$ as a function of $M_{inv}$.}
\label{fig:Y01}
\end{figure}

\begin{figure}
\centering
\includegraphics[width=0.8\textwidth]{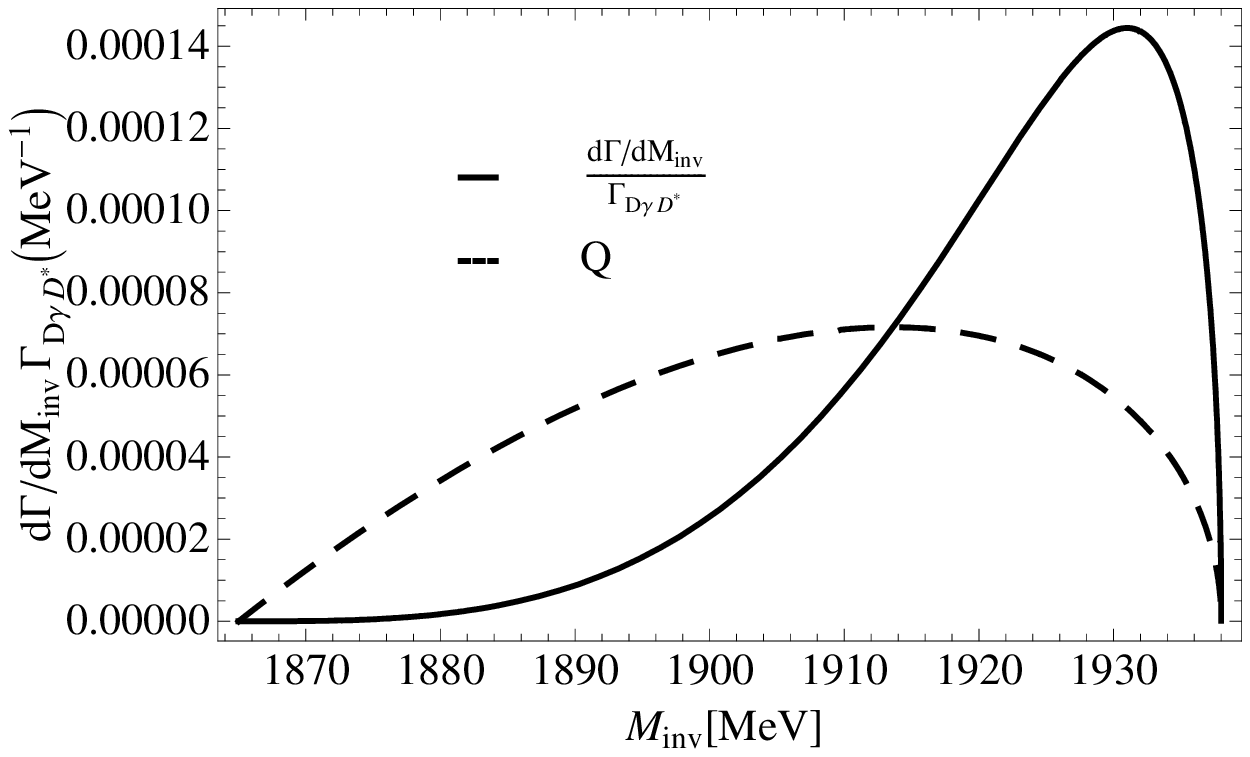}
\caption{The $Y_p(3945)\to D^{*0}\bar{D}^0\gamma$: $d\Gamma/dM_{inv}\Gamma_{D^*D\gamma}$ and $Q$ as a function of $M_{inv}$.}
\label{fig:Y01d}
\end{figure}

 This state has zero width, and here we show the difference between $d\Gamma_R/dM_{inv}$ and $Q$ in the case of $Y_p(3945)\to D^{*+}D^{-}\gamma$ in Fig. \ref{fig:Y01} to see the effect of the inclusion of the $\bar{D}^*$ propagator in Eq. (\ref{eq:tdos}). As one can see, the shapes can be clearly distinguished. Also, in Fig. \ref{fig:Y01d} we show the curves for $d\Gamma_R/dM_{inv}\Gamma_{D^*D\gamma}$ for the case of the neutral charm mesons in the final state.

\subsection{The Z(3930)}
\begin{figure}
\centering
\includegraphics[width=1\textwidth]{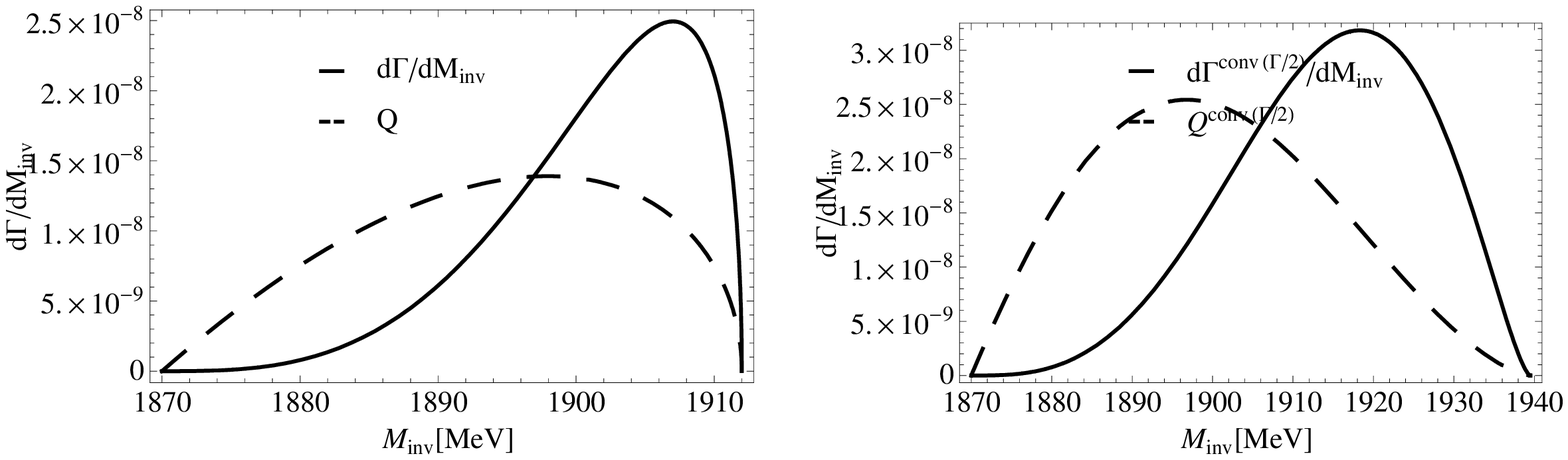}
\caption{The $Z(3930)\to D^{*+}D^-\gamma$: a) $d\Gamma/dM_{inv}$ and $Q$ as a function of $M_{inv}$. b) $d\Gamma^{\mathrm{conv}}/dM_{inv}$ and $Q^{\mathrm{conv}}$.}
\label{fig:Z02}
\end{figure}

 This state has a larger width compared with the Y(3940) and $Y_p(3945)$ states of $55$ MeV, and for this reason the picture here is very different than in those cases when one takes into account this width. Thus, one can see a big difference between $d\Gamma_R/dM_{inv}$ and $d\Gamma^{\mathrm{conv}}_R/dM_{inv}$, $Q$ and $Q^{\mathrm{conv}}$, as shown in Fig. \ref{fig:Z02}. The relatively large width of the resonance taken ($55$ MeV) is responsible for the different shapes compared to Fig. \ref{fig:Z02} a). Similar results are obtained for $d\Gamma/dM_{inv}\Gamma_{D^*D\gamma}$ for decay into $D^{*0}\bar{D}^0\gamma$.


\subsection{The $Y_p(3912)$}
\begin{figure}
\centering
\includegraphics[width=0.95\textwidth]{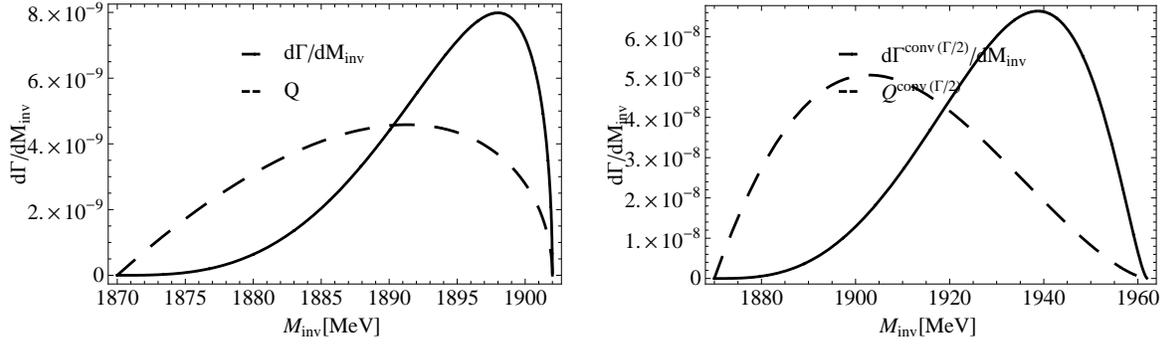}
\caption{The $Y_p(3912)\to D^{*+}D^-\gamma$: a) $d\Gamma/dM_{inv}$ and $Q$ as a function of $M_{inv}$. b) $d\Gamma^{\mathrm{conv}}/dM_{inv}$ and $Q^{\mathrm{conv}}$.}
\label{fig:Y12}
\end{figure}

 This case is very similar to that of the Z(3930). The shapes of $d\Gamma_R/dM_{inv}$ and $Q$ are very different (also for $d\Gamma^{\mathrm{conv}}_R/dM_{inv}$ and $Q^{\mathrm{conv}}$) as one can see in Fig. \ref{fig:Y12}. Now the width is considerably larger compared to that in the previous cases, since $\Gamma=120$ MeV. Similar results are obtained for the case of $Y_p(3912)\to D^{*0}\bar{D}^0\gamma$.

\subsection{The X(4160)}
\begin{figure}
\centering
\includegraphics[width=1\textwidth]{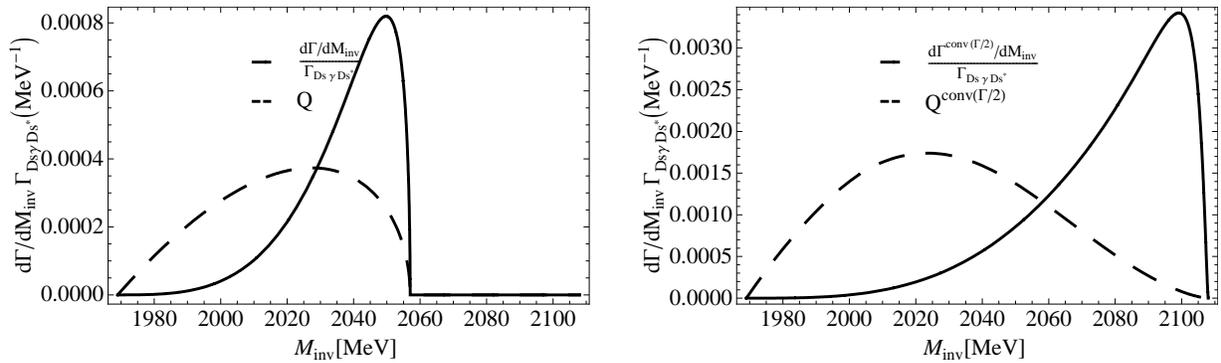}
\caption{The $X(4160)\to D^{*+}_sD^-_s\gamma$: a) $d\Gamma_R/dM_{inv}\Gamma_{D_s^*D_s\gamma}$ and $Q$ as a function of $M_{inv}$. b) $d\Gamma^{\mathrm{conv(\Gamma/2)}}_R/dM_{inv}\Gamma_{D_s^*D_s\gamma}$ and $Q^{\mathrm{conv}}$.}
\label{fig:X02}
\end{figure}

 In this case the isospin
factor is $\mathrm{F_I}=1$ rather than $1/\sqrt{2}$.  The formula is the same as before removing
a factor $1/2$ in Eq. (\ref{eq:tdos}). Once again we do not have the experimental decay rate
for the radiative decay of $D_s^{*-}$ and we plot the results for 
Eq. (\ref{eq:gammarel}) in
Fig. \ref{fig:X02}. In this case the decay into $D^{*+} D^- \gamma$ is also
possible. However, the coupling to $D^* \bar{D}^*$ of this resonance 
(also assumed to be a
$D^{*+}_s \bar{D}_s^{*-}$ molecule in \cite{liuke}  and \cite{gutsche}) is
found small in \cite{xyz}, of the order of $17$ times smaller, hence the rate for
this channel should be drastically smaller. In order to test the $D^{*+}D^{*-}$ component of this molecule, the allowed strong
decay into  $D^* \bar{D}^*$ is preferable. This
latter measurement is a more efficient tool to get the strength of this coupling and
compare with the theoretical predictions.

\begin{table}[htb]
\begin{center}
\begin{tabular}{cccccc}
\hline
State&Decay&$\Gamma$\,[keV]\T\B&$\Gamma/\Gamma_{D^{*-}_{(s)}\to D^-_{(s)}\gamma}$&$\Gamma^{\mathrm{conv(\Gamma/2)}}$\,[keV]&$\Gamma^{\mathrm{conv(\Gamma/2)}}/\Gamma_{D^{*-}_{(s)}\to D^-_{(s)}\gamma}$\\
\hline
\hline
Y(3940)&$D^{*+}D^-\gamma$&$2.7\times 10^{-3}$&$1.8\times 10^{-3}$&\T\B$2.9\times 10^{-3}$&$1.9\times 10^{-3}$\\
$Y_p(3945)$&$D^{*+}D^-\gamma$&$3.1\times 10^{-3}$&$2.0\times 10^{-3}$&$-$ &$-$\T\B\\
Z(3930)&$D^{*+}D^-\gamma$&$4.1\times 10^{-4}$&$2.6\times 10^{-4}$&$1.0\times 10^{-3}$&$6.7\times 10^{-4}$\T\B\\
$Y_p(3912)$&$D^{*+}D^-\gamma$&$1.0\times 10^{-4}$&$6.7\times 10^{-5}$&$2.7\times 10^{-3}$&$1.8\times 10^{-3}$\T\B\\
X(4160)&$D_s^{*+}D^-_s\gamma$&$<39.9$&$2.3\times10^{-2}$&$<2.4\times 10^2$&$0.14$\T\B\\
\hline
\end{tabular}
\end{center}
\caption{Decay of the XYZ resonances into $D^{*+}D^-\gamma$ and $D_{s}^{*+}D_{s}^-\gamma$.}
\label{tab:decay1}
\end{table} 

\begin{table}[htb]
\begin{center}
\begin{tabular}{cccccc}
\hline
State&Decay&$\Gamma$\,[keV]\T\B&$\Gamma/\Gamma_{\bar{D}^{*0}\to \bar{D}^0\gamma}$&$\Gamma^{\mathrm{conv(\Gamma/2)}}$\,[keV]&$\Gamma^{\mathrm{conv(\Gamma/2)}}/\Gamma_{\bar{D}^{*0}\to \bar{D}^0\gamma}$\\
\hline
\hline
Y(3940)&$D^{*0}\bar{D}^0\gamma$&$<2.6$&$3.2\times 10^{-3}$&\T\B$<2.7$&$3.4\times 10^{-3}$\\
$Y_p(3945)$&$D^{*0}\bar{D}^0\gamma$&$<2.9$&$3.6\times 10^{-3}$&$-$&$-$\T\B\\
Z(3930)&$D^{*0}\bar{D}^0\gamma$&$<0.48$&$6.0\times 10^{-4}$&$<1.0$&$1.3\times 10^{-3}$\T\B\\
$Y_p(3912)$&$D^{*0}\bar{D}^0\gamma$&$<0.15$&$1.9\times 10^{-4}$&$<2.4$&$3.0\times 10^{-3}$\T\B\\
\hline
\end{tabular}
\end{center}
\caption{Decay of the XYZ resonances into $D^{*0}\bar{D}^0\gamma$.}
\label{tab:decay2}
\end{table} 



In Tables \ref{tab:decay1} and \ref{tab:decay2} we show integrated values for $\Gamma(R\to PV\gamma)$ and also rates of $\Gamma(R\to PV\gamma)$ with respect to $\Gamma(D^*_{(s)}\to D_{(s)}\gamma)$. In the case of the decays of the resonance into $D^{*0}\bar{D}^0\gamma$, $D^{*+}_s D^-_s\gamma$, $D^{*0}\bar{D}^0\pi^0$ and $D^{*+}_s D^-_s\pi^0$, we compute $g_{PV\gamma}$ in Eq. (\ref{eq:radwidth}) taking $\Gamma(D^{*0})<2.1$ MeV and $\Gamma(D^{*+}_s)<1.9$ MeV. We show in Table \ref{tab:decay1}, the integrated values for $\Gamma(R\to D^{*+}D^-\gamma)$ which are very small, of the order of $10^{-1}-1$ eV if one does not consider the convolution of the $d\Gamma/dM_{inv}$ distribution. However, when one considers the width of the XYZ resonances given in Table \ref{tab:exp}, these values become bigger (about one order of magnitude in some cases).

In the case of the X(4160) we can only put a boundary for the $\Gamma(X\to D^{*+}_s D_s^-\gamma)$, which is $39.9$ KeV, but we give rates of $\Gamma(X\to D^{*+}_s D_s^-\gamma)$ respect to $\Gamma(D^{*-}_s \to D^-_s\gamma)$ in Table \ref{tab:decay1}. For this observable we get a value of $2.3\times 10^{-2}$ and $0.14$ before and after convolution respectively. When the final state contains neutral charm mesons, we give both amplitudes and rates which can be seen in Table \ref{tab:decay2}. In Table \ref{tab:decay2}, we see that $\Gamma/\Gamma_{\bar{D}^{*0}\to \bar{D}^0\gamma}$ is of the order of $10^{-4}-10^{-3}$ for all the states before the convoluting $d\Gamma_R/dM_{inv}$ and becomes larger when one convolutes this function.

\section{Summary}

We have presented results for decay of the heavy dynamically generated states
from the vector-vector interaction, with hidden charm, into  $D^*$ and $\bar{D}
\gamma$, or $D^*_s$ and $\bar{D}_s \gamma$. We find a very
distinctive shape in the $\bar{D} \gamma$ and $\bar{D}_s \gamma$ invariant mass
distributions, which is
peculiar to the molecular nature of these states as basically bound states of
two charmed vector mesons. It was suggested in \cite{xyz} that some of these
states correspond to some of the X,Y,Z states found at the Belle and Babar
facilities.  We hope the findings of the present paper stimulate  experimental
work in this direction to further learn about the nature of the X,Y,Z
resonances.

\section*{Acknowledgments}
This work is partly supported by DGICYT contract number
FIS2006-03438. We acknowledge the support of the European Community-Research
Infrastructure Integrating Activity
"Study of Strongly Interacting Matter" (acronym HadronPhysics2, Grant Agreement
n. 227431) under the Seventh Framework Programme of EU.


\begin{thebibliography}{99}



\bibitem{review}
  J.~A.~Oller, E.~Oset and A.~Ramos,
  Prog.\ Part.\ Nucl.\ Phys.\  {\bf 45}, 157 (2000)

\bibitem{hidden1}
  M.~Bando, T.~Kugo, S.~Uehara, K.~Yamawaki and T.~Yanagida,
  Phys.\ Rev.\ Lett.\  {\bf 54}, 1215 (1985).


\bibitem{hidden2}
  M.~Bando, T.~Kugo and K.~Yamawaki,
  Phys.\ Rept.\  {\bf 164}, 217 (1988).

\bibitem{hidden3}
  M.~Harada and K.~Yamawaki,
  Phys.\ Rept.\  {\bf 381}, 1 (2003)

\bibitem{Bernard:1988db}
  V.~Bernard and U.~G.~Meissner,
  Nucl.\ Phys.\  A {\bf 489}, 647 (1988).

\bibitem{raquel}
  R.~Molina, D.~Nicmorus and E.~Oset,
  Phys.\ Rev.\  D {\bf 78}, 114018 (2008)
  
\bibitem{geng}
  L.~S.~Geng and E.~Oset,
  Phys.\ Rev.\  D {\bf 79}, 074009 (2009)
  
\bibitem{ollerulf}
  J.~A.~Oller and U.~G.~Meissner,
  Phys.\ Lett.\  B {\bf 500}, 263 (2001)

\bibitem{klempt}
  E.~Klempt and A.~Zaitsev,
  Phys.\ Rept.\  {\bf 454}, 1 (2007)


\bibitem{crede}
  V.~Crede and C.~A.~Meyer,
  Prog.\ Part.\ Nucl.\ Phys.\  {\bf 63}, 74 (2009)





\bibitem{yamagata}
  H.~Nagahiro, J.~Yamagata-Sekihara, E.~Oset, S.~Hirenzaki and R. Molina,
  Phys.\ Rev.\  D {\bf 79}, 114023 (2009)


\bibitem{chinacola}
  A.~Martinez Torres, L.~S.~Geng, L.~R.~Dai, B.~X.~Sun, E.~Oset and B.~S.~Zou,
  Phys.\ Lett.\  B {\bf 680}, 310 (2009)

\bibitem{chinavalgerman}  
  L.~S.~Geng, F.~K.~Guo, C.~Hanhart, R.~Molina, E.~Oset and B.~S.~Zou,
  arXiv:0910.5192 [hep-ph].
\bibitem{BranzGeng}
  T.~Branz, L.~S.~Geng and E.~Oset,
  arXiv:0911.0206 [hep-ph].

\bibitem{raquelnaga}
  R.~Molina, H.~Nagahiro, A.~Hosaka and E.~Oset,
  Phys.\ Rev.\  D {\bf 80}, 014025 (2009)



\bibitem{pdg}
  C.~Amsler {\it et al.}  [Particle Data Group],
  Phys.\ Lett.\  B {\bf 667}, 1 (2008).



\bibitem{xyz}
  R.~Molina and E.~Oset,
  Phys.\ Rev.\  D {\bf 80}, 114013 (2009)


\bibitem{shilinzhu}
  X.~Liu, Z.~G.~Luo, Y.~R.~Liu and S.~L.~Zhu,
  Eur.\ Phys.\ J.\  C {\bf 61}, 411 (2009)

\bibitem{gutsche}
  T.~Branz, T.~Gutsche and V.~E.~Lyubovitskij,
  Phys.\ Rev.\  D {\bf 80}, 054019 (2009)


\bibitem{liuke}
  X.~Liu and H.~W.~Ke,
  Phys.\ Rev.\  D {\bf 80}, 034009 (2009)














\end{thebibliography}
\end{document}